\documentclass[%
 reprint,
superscriptaddress,
 amsmath,amssymb,
 aps,
pra
]{revtex4-2}

\usepackage{xcolor}
\usepackage{graphicx}
\usepackage{dcolumn}
\usepackage{bm}
\usepackage[normalem]{ulem}

\begin{document}

\title{Twisted hyperbolic van der Waals crystals for chip-scale full Stokes mid-infrared polarization detection}

\author{Nihar Ranjan Sahoo}
\thanks{These two authors contributed equally}
\author{S.S. Jatin Prasath}
\thanks{These two authors contributed equally}
\author{Brijesh Kumar}
\affiliation{Laboratory of Optics of Quantum Materials, Physics Department, Indian Institute of Technology Bombay, Mumbai - 400076, India}
\author{Anshuman Kumar}
\email{anshuman.kumar@iitb.ac.in}
\affiliation{Laboratory of Optics of Quantum Materials, Physics Department, Indian Institute of Technology Bombay, Mumbai - 400076, India}
\affiliation{Centre of Excellence in Quantum Information, Computation, Science and Technology, Indian Institute of Technology Bombay, Powai, Mumbai- 400076, India} 
\affiliation{Center for Semiconductor Technologies (SemiX), Indian Institute of Technology Bombay, Mumbai - 400076, India}

\begin{abstract}
Investigating the polarization properties of light in the mid-infrared (mid-IR) spectrum is crucial for molecular sensing, biomedical diagnostics, and IR imaging system technologies. Traditional methods, limited by bulky size and complicated fabrication process, utilize large rotating optics for full Stokes polarization detection, impeding miniaturization and accuracy.Naturally occurring hyperbolic van der Waals (vdW) material based devices can address these challenges due to their lithography-free fabrication, ease of integration with chip-scale platforms and room-temperature operation. This study designs a chip-integrated polarimeter by performing multi-objective optimization for efficient exploration of the design parameter space. The spatial division measurement scheme used incorporates six precisely designed linear and circular polarization filters, achieving high extinction ratios exceeding 30 dB and transmittance surpassing 50\%, with fabrication tolerance of film thickness up to 100 nm. The proposed device represents a significant advancement in polarimetric detection, providing a compact, cost-effective solution and opens new avenues for on-chip mid-IR polarimetric detection in next-generation ultra-compact optical systems.
\end{abstract}

\maketitle

\section*{Introduction}
Precisely manipulating and detecting the polarization state of light holds significant implications across scientific and technological domains. This significance is especially notable in the mid-IR spectrum, where precise characterization of the polarization state of light contributes to advancements in molecular sensing\cite{Martnez2016}, biomedical diagnostics\cite{Oh2018}, thermal imaging\cite{Park2018}, and IR imaging system technologies\cite{Tong2020}. Unfortunately, conventional mid-IR polarization detection methods rely on rotating optical components, which present challenges in device integration, miniaturization, speed, and accuracy.\cite{wu2008optical,schaefer2007measuring} Hence, there's a critical need to develop compact and cost-effective mid-IR polarimetric detection systems. Recent advancements in monolithically integrated polarimetric imaging systems have showcased the potential by utilizing micropolarization filters atop silicon photodetectors, simplifying on-chip integration and enhancing reliability.\cite{tokuda2009polarisation,zhao2009thin} However, achieving comprehensive polarization state detection involves the incorporation of multiple linear polarization (LP) and circular polarization (CP) filters. Current polarization filter technologies, including birefringent materials\cite{andreou2002polarization}, thin-film polarizers\cite{zhao2009thin,shopsowitz2010free,gruev2007fabrication}, metallic nanowires\cite{wang2007high,gruev2010ccd}, and organic materials\cite{coleman2003polarization,sánchez2014circularly}, exhibit limitations such as scalability issues, structural instability, high absorption in mid-IR regions, and the inability to achieve complete polarization state measurement. Hence, accomplishing effective mid-IR polarimetric detection remains a challenge, demanding immediate efforts towards miniaturization and the development of cost-effective solutions with on-chip integrated technology.

Hyperbolicity\cite{Low2016}, an extreme form of birefringence showcased when certain crystal axes possess a negative dielectric permittivity while others remain positive, addresses anisotropy and substantially reduces device dimensions by a few orders of magnitude.\cite{ren2012giant} This property forms the basis for remarkable optical phenomena like hyperlensing\cite{liu2007far},canalization\cite{correas2017plasmon}, enhanced thermal radiation\cite{shi2017enhanced}, negative refraction\cite{lin2017all}. 
Conventionally, achieving these properties relied on metamaterials, requiring complex fabrication techniques involving sub-wavelength periodic features, compromising device performance.\cite{Guo2020,wadsworth2011broadband,wu2014spectrally} However, the emergence of natural hyperbolic van der Waals (vdW) crystals, bypassing expensive micro- \& nanofabrication efforts, has unveiled new possibilities for birefringent optical components and polarization-dependent photonics.\cite{Guo2020,korzeb2015compendium} Examples of such natural hyperbolic vdW crystals include black phosphorus (BP)\cite{Yang2017,PhysRevB.96.235403}, h-BN\cite{Caldwell2014,li2018infrared}, $\alpha$-V$_2$O$_5$\cite{TaboadaGutirrez2020}, $\alpha$-MoO$_3$\cite{Zheng2019,Ma2018,lvarezPrez2020}, $\beta$-Ga\textsubscript{2}O\textsubscript{3}\cite{passler2022hyperbolic}. 

\begin{figure*}
\centering
\includegraphics[width=\textwidth]{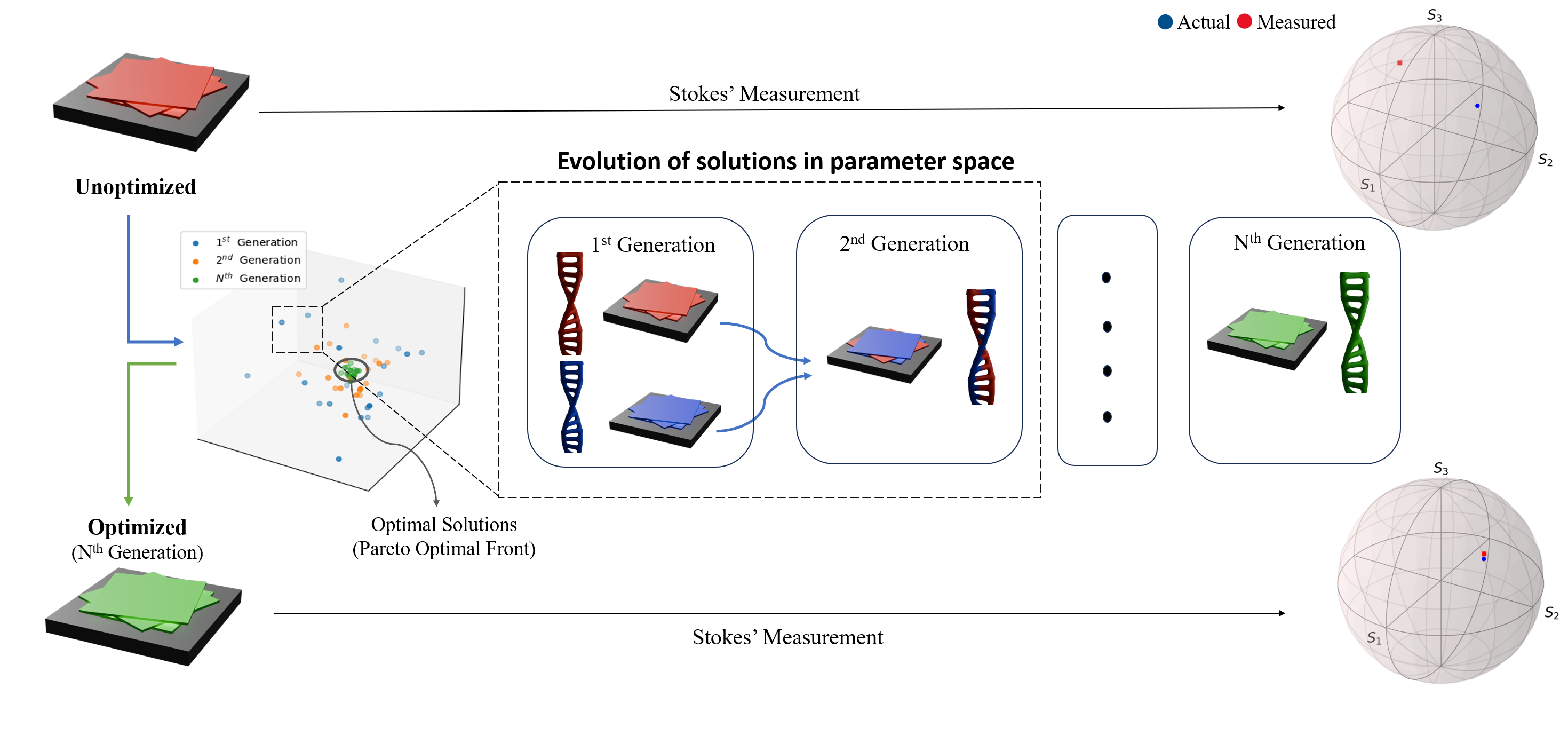}
\caption{\textbf{Design process of Stokes polarimeter:} The first-generation unoptimized filter is unable to produce the desired characteristic, resulting in an inaccurate mapping of transmitted intensity to the Stokes parameter. As a result, the measured Stokes parameter deviates significantly from its actual value. To address this problem, NSGA-II is utilized to optimize the filter and ensure a faithful mapping. The first-generation filters with a range of film stacks of varying thickness and twist angles are represented by blue dots in parameter space (red and blue colours film stacks). In the second generation, orange dots in parameter space, a crossover set of parameters from the first-generation was used, resulting in a mixture of red and blue colour films in the second generation filters. Finally, after several generations, the optimized filter was achieved and is represented by a green colour.}
\end{figure*}

There are efforts to integrate vdW crystals into on-chip photonics, which has generated interest in developing optical components.\cite{grigorenko2012graphene} These materials can be exfoliated into desired thicknesses and transferable onto diverse substrates, offering opportunities for creating 2D heterostructures.\cite{geim2013van} While materials like hBN have been extensively studied in the context of hyperbolic phonon polaritons (PhPs), their in-plane anisotropy is not naturally supported due to crystal lattice symmetry.\cite{li2018infrared} Conversely, graphene \textcolor{black}{ribbons} and BP showcase various degrees of in-plane anisotropy\cite{cheng2013dynamically,kotov2017enhanced}, yet the efficiency due substrate impact\cite{kotov2017enhanced} and plasmonic losses\cite{Low2016} hinder their practical device application. In contrast, the highly anisotropic $\alpha$-MoO\textsubscript{3}, a biaxial vdW crystal in the mid-IR, exhibits naturally occurring in-plane hyperbolicity, offering a platform with low-loss optical phonons for a higher degree of polarization control without the need for lithography techniques.\cite{dixit2021mid} $\alpha$-MoO$_3$ have orthorhombic unit cell structure composed of distorted MoO\textsubscript{6} octahedra, exhibits three different lattice constants, notably displaying a substantial 7\% difference in its in-plane lattice constants, resulting in robust in-plane anisotropy.\cite{Ma2018} \textcolor{black}{The dielectric permittivity tensor of $\alpha$-MoO\textsubscript{3} is accurately described by the Lorentz oscillator model, which accurately describes the frequency-dependent optical responses due to optical phonons.\cite{alvarez2020infrared} At specific frequencies corresponding to the transverse optical (TO) phonons, the real part of the dielectric permittivity becomes negative along one crystallographic direction while remaining positive in the orthogonal directions, creating three distinct Reststrahlen (RB) Bands for the case of $\alpha$-MoO\textsubscript{3}. This model's validity is supported by first-principles-based calculations that consider contributions from electronic and phononic interactions along different crystallographic axes \cite{tong2021first,ding2012structural,inzani2016van}. It is important to include these IR active modes to ensure that the permittivity model remains accurate across the relevant length scales considered for the polarimeter.} Further, the fabrication of thin films of $\alpha$-MoO\textsubscript{3} can be achieved through cost-effective methods like physical vapor deposition\cite{sahoo2023polaritons},  providing thickness variations from a few hundred nanometers to microns, enables the generation of diverse thicknesses of these vdW single crystals. This flexibility complements the intrinsic in-plane hyperbolic anisotropy of $\alpha$-MoO\textsubscript{3}, holding promise for highly efficient and compact mid-IR optical devices.

This work presents a novel, chip-integrated polarimeter for accurately measuring the complete state of mid-IR light by harnessing hyperbolic properties of $\alpha$-MoO\textsubscript{3} on a silicon substrate, operating in the transmittance mode. Unlike traditional bulky systems, the proposed device utilizes unpatterned thin films of vdW crystals with variable thickness and twist angles, making it a lithography-free alternative for designing an efficient polarimeter device. We considered several LP and CP filters on the proposed single-chip device to measure the full-Stokes parameters of the arbitrarily polarized incident light. The optimal design of the filters is crucial for obtaining an accurate measurement of stokes parameter. As shown in Fig. 1 an unoptimized filter in the polarimeter will not produce a faithful mapping of measured intensities to the actual Stokes parameters. Thus to efficiently explore the design space and obtain the optimal solution, multiobjective optimization was performed using NSGA-II\cite{deb2002fast}. The scatter plot in Fig. 1 shows NSGA-II's search for optimal solution in parameter space where a set of random parameters are weighted using the objective functions, and parameters with the lowest values of the objective function are selected and crossed to create future generations, which ultimately converge to a set of optimal solutions (Pareto optimal front) from which the optimal solution is selected using performance metrics. We find that for any designed filter, the performance metric extinction ratio (ER) is above 30 dB with a minimum 50\% transmittance in the operating frequency range. These results are notable compared to other flat optics metasurfaces-based mid-IR CP filters\cite{gansel2009gold,frank2013large} and polarimeter devices\cite{bai2019chip,jung2018polarimetry,bai2021highly} which are built using lithographic techniques, hence, a more suitable candidate for integration into mid-IR chip-scale devices, photodetectors, imaging sensors.

\section*{Methods}

\begin{figure*}
\centering
\includegraphics[width=\textwidth]{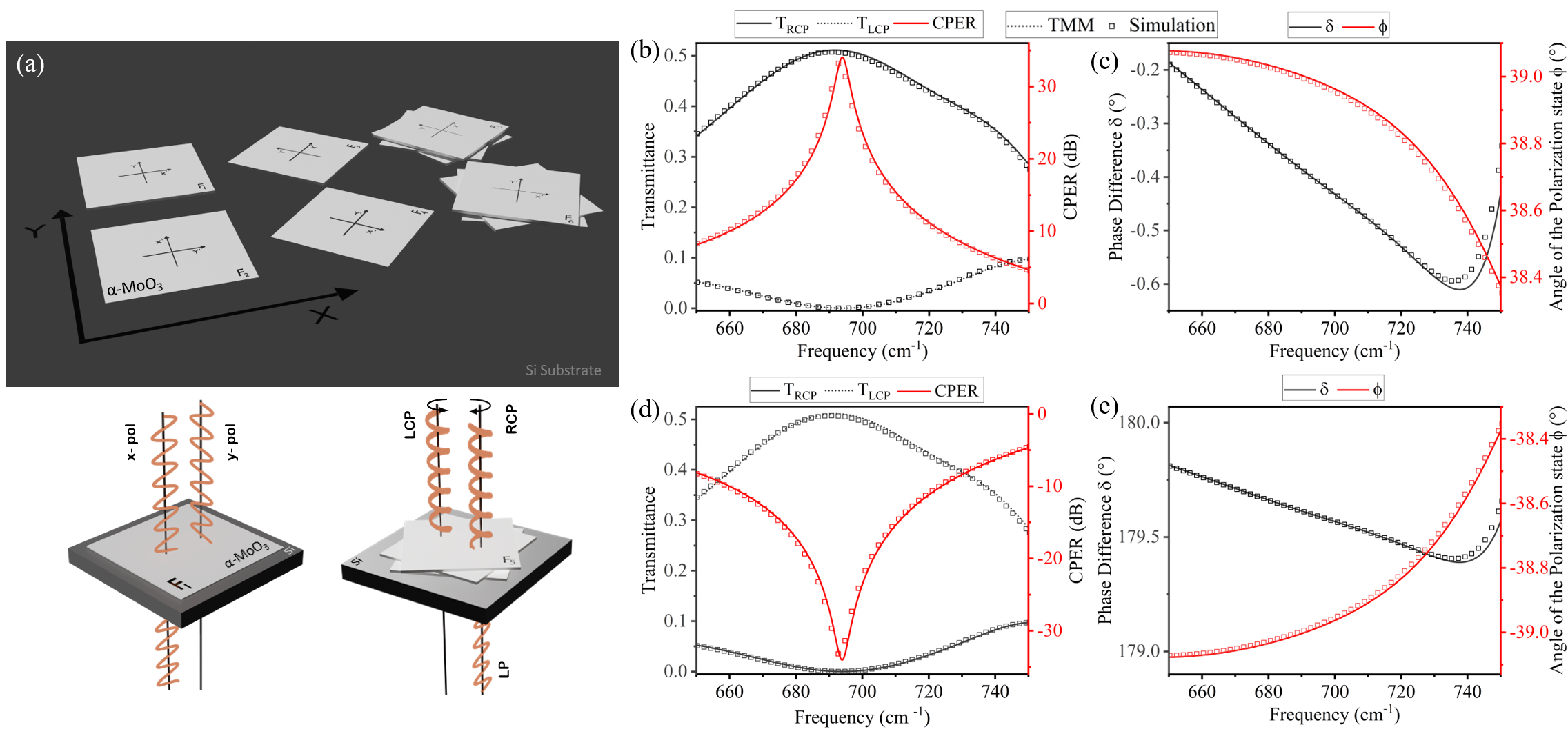}
    \caption{ \textbf{Frequency response of designed circular polarizers:} (a)  Schematic illustration of a full Stokes polarimeter with six $\alpha$-MoO\textsubscript{3}-based filters (F1-F6) on a Si substrate. Each filter comprises multiple $\alpha$-MoO\textsubscript{3} layers with varying thickness and twist angles. F1 to F4 are LP filters, F5 and F6 are RCP and LCP filters. Inset details F1's x-polarized light transmission while blocking y-polarized light and F5's conversion of RCP light to linearly polarized, blocking LCP light. Figure (b) and (d) shows transmittance due to incident RCP and LCP light as solid and dashed lines, where CPER represents in solid red line for filters F5 and F6, respectively.  The phase difference ($\delta$) between $x$- and $y$- components, and angle of the polarization state ($\phi$) of transmitted light from filter $F_5$ due to incident RCP light,(c) and from F6 due to LCP light (e) are shown in black and red solid lines, respectively. Further, Scatter plots and line plots in (b–e) represent optical parameters obtained from numerical simulations and TMM. }
\end{figure*}

The designed polarimeter is based on spatial division measurement scheme. This scheme comprises six polarization filters denoted as $F_1$ to $F_6$, as depicted in Fig. 2(a). Among these filters, $F_1$ to $F_4$ are LPs. Each LP is oriented to allow the transmission of light based on their polarization state with respect to the x-axis: at 0$^{\circ}$ ($F_1$), 90$^{\circ}$ ($F_2$), 45$^{\circ}$ ($F_3$), and -45$^{\circ}$ ($F_4$). These orientations are crucial as they selectively permit light aligned with these angles to pass through while completely blocking their orthogonal polarization states. Filters $F_5$ and $F_6$ are CP filters which allow either right circularly polarized (RCP) or left circularly polarized (LCP) light to pass through. Filter $F_5$ is designed to convert incoming  RCP light into linearly polarized light while blocking LCP light completely. (Fig. 2(a) inset) Conversely, in filter $F_6$, incident LCP light transforms into linearly polarized light, while RCP light is completely blocked. 

By measuring the transmitted intensity of the individual filters for a given incident light, the Stokes parameter\cite{stokes1851composition} are obtained as

\begin{equation}
\begin{array}{l}
    S_{0} = I_{o}\\
    S_{1} = I_{0^{o}} - I_{90^{o}}\\
    S_{2} = I_{45^{o}} - I_{-45^{o}}\\
    S_{3} =I_{RCP} - I_{LCP}
\end{array}
\end{equation}

where, $I_o$ is the intensity of the incident light, $I_{0^o}$, $I_{90^o}$, $I_{45^o}$, $I_{-45^o}$, $I_{RCP}$ and $I_{LCP}$ are the intensities of the light transmitted through ideal polarization filters $F_1$ to $F_6$ respectively. By normalizing each Stokes parameter with $S_0$, we can reduce the number of parameters to 3 without any loss of generality.\par       

\begin{table}
\centering
\setlength{\tabcolsep}{10pt}
\renewcommand{\arraystretch}{1.5}
\begin{tabular}{|m{1.5cm} | m{1.5 cm}| m{1.7cm}| }
\hline
Components & Thickness (nm)  & Twist angle (deg)  \\
\hline
$F_1$ & 5000 & 0 \\

$F_2$ & 5000 & 90 \\

$F_3$ & 5000 & 45 \\

$F_4$ & 5000 & -45 \\

$F_5$ & 557 & 35 \\

 & 2100 & 110 \\

 & 3487 & 129 \\

$F_6$ & 557 & -35 \\

 & 2100 & -110 \\

 & 3487 & -129 \\

\hline
\end{tabular}
\caption{\textbf{Design parameters for $\alpha$-MoO$\textsubscript{3}$ based polarization filters.} Here, $F_1$ to $F_4$ represent LP filters, $F_5$ and $F_6$ denote RCP and LCP filters. The LP filters consist of a single layer with different twist angles, whereas the CP filters are constructed with three layers, each having distinct twist angles. Note that these are the mean values, whereas the tolerance analysis is provided later in the manuscript.} 
\label{Table:1}
\end{table}

It is necessary to have LP and CP filters with high ER to accurately determine the Stokes parameter via the proposed polarimeter design. Fabricating LPs with high ER in mid-IR is relatively simple\cite{sahoo2022high}. However, obtaining CP filters with high ER using $\alpha$-MoO$\textsubscript{3}$  thin films is quite challenging due to the requirement of a larger design parameter space for film stack. To optimally explore the design parameter space and to make the CP filters based on $\alpha$-MoO$\textsubscript{3}$ thin films, we approached the task as an optimization problem. The design specification for the CP filter comprises multiple parameters, including the total number of $\alpha$-MoO$\textsubscript{3}$ films in the stack, their individual thickness and twist angle,  and the operating frequency at which the highest achievable ER  can be attained. In our approach, we consider the thickness and in-plane twisting of each layer, along with the operating frequency, as the optimization parameters, while the total number of layers in the film stack serves as a hyper-parameter. For designing $F_5$ (RCP filter), we minimised the following objective functions \textbf{\{$-  T_{RCP}$, $T_{LCP}$, $-CPER$\}}. Here, $T_{RCP}$ and $T_{LCP}$ denote the transmittance due to incident RCP and LCP light, respectively, along with circular polarization extinction ratio ($CPER$), which is defined as  $10log(T_{RCP}/T_{LCP})$. NSGA-II algorithm as provided by Pymoo\cite{pymoo}, was used to perform constrained optimization of the mentioned objective functions. The thickness of individual films was constrained to be in the range 50-5000 nm, the twisting to be in the range 0$^{\circ}$-170$^{\circ}$ and the operating frequency was constrained in $\alpha$-MoO$\textsubscript{3}$ RS band 1\cite{dixit2021mid} which is in the range 540 cm$\textsuperscript{-1}$ - 820 cm$\textsuperscript{-1}$. The same set of constraints were used for the design of all the other filters. The evaluation criterion for  $F_5$ was that CPER must be greater than 20 dB. The design parameters for $F_6$ (LCP filter) are identical to those of F5, except for the orientation of the layers, which is the negative of the twist angle used for $F_5$'s films. This ensures a complementary arrangement between the orientations of the films in $F_5$ and $F_6$, addressing the requirements for achieving LCP filter.

LP filters with high ER can be obtained using a single thin $\alpha$-MoO$\textsubscript{3}$  film with appropriate thickness.\cite{sahoo2022high} The optimization parameter for $F_1$ is the film's thickness and the operating frequency, which has fixed to be the same as that of the CP filter. To obtain the optimal film thickness,  the following parameters \textbf{\{$-T_{x}$, $T_{y}$, $-LPER$\}} were minimized, where $T_{x}$ and $T_{y}$ represent the transmittance due to incident light polarized along $x$-(0$^{\circ}$) and $y$-(90$^{\circ}$) axis, respectively. The evaluation criteria for designing $F_1$ was that the linear polarized extinction ratio (LPER), which is defined as $10log(T_{x}/T_{y})$, must be greater than 30 dB. The subsequent LP filters, $F_2$ to $F_4$, were obtained by rotating $F_1$ by 90$^{\circ}$, 45$^{\circ}$, and -45$^{\circ}$, respectively, preserving the same film thickness while altering the orientation. Table 1 shows the obtained design parameters of polarization filters $F_1$-$F_6$.

\section*{Results and discussion}

\begin{figure}[h]
\centering
\includegraphics[width=\columnwidth]{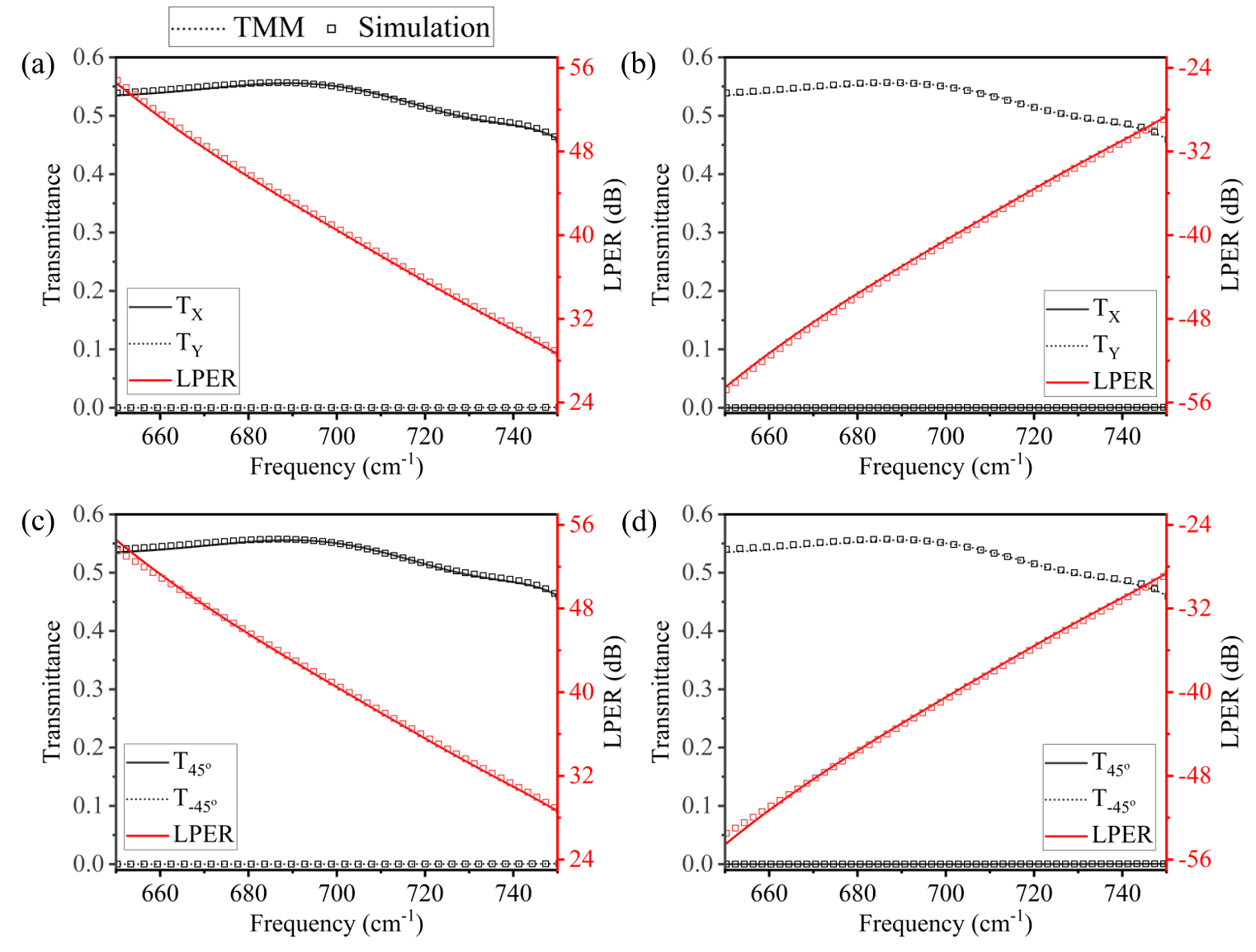}
\caption{\textbf{Frequency response of designed linear polarizers:} (a-d) corresponds to the transmittance spectra and LPER in black and red solid lines, respectively, for LP filters F1 to F4. The incident linearly polarized light is oriented along $x$- axis (0$^{\circ}$) and $y$- axis (90$^{\circ}$) in (a-b), and at 45$^{\circ}$ and -45$^{\circ}$ in (c-d), respectively. The mean thickness of the $\alpha$-MoO$\textsubscript{3}$ layer is fixed at 5$\mu$m, and with twist angle 0$^{\circ}$, 90$^{\circ}$, 45$^{\circ}$ and -45$^{\circ}$, respectively, for filter $F_1$-$F_4$. Further, Scatter plots and line plots represent optical parameters obtained from numerical simulations and TMM.}
\end{figure}

\begin{figure*}
\centering
\includegraphics[width=\textwidth]{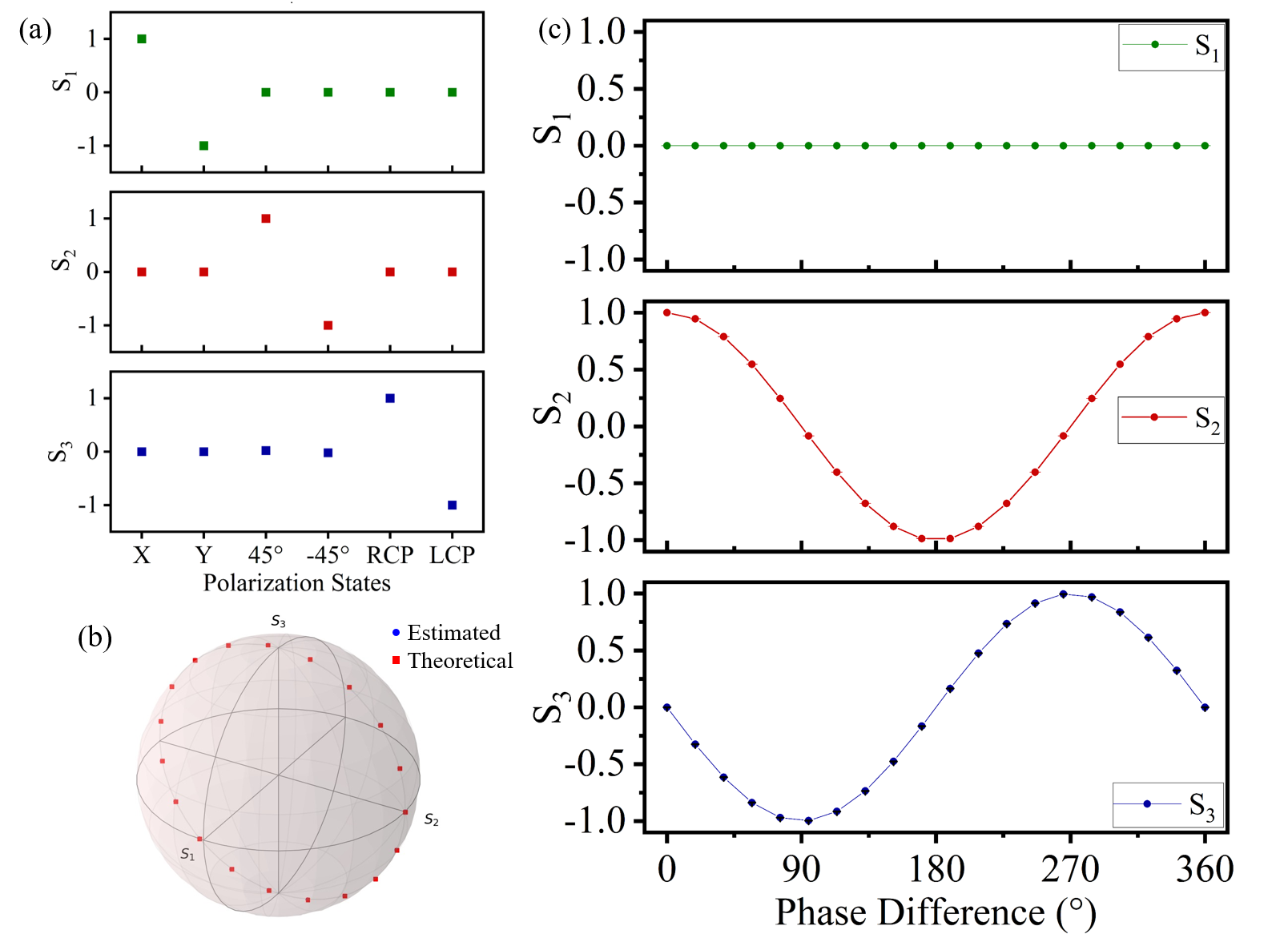}
\caption{ \textbf{Characterization of Stokes parameter:} (a) Measured Stokes parameters for some common states of polarization, including linearly polarized light along the $x$-axis, $y$-axis, 45$^{\circ}$, -45$^{\circ}$, RCP and LCP, (b) Stokes parameters for incident elliptically polarized light with varying phase difference on a Poincaré sphere. The blue dots represent the estimated values, while the theoretical values are indicated by red dots. Here both the points are overlapping, representing a negligible error between the estimated and theoretical values. (c) All three Stokes parameters of the incident elliptically polarized light are plotted as a function of phase difference. The incident elliptically polarized light possesses equal $x$- and $y$-components of the electric field, with the phase difference ranging from 0$^{\circ}$ to 360$^{\circ}$.}
\end{figure*}

The frequency response of the designed polarization filters is studied to obtain operating bandwidth. Fig. 2(b) shows the variations in $T_{RCP}$, $T_{LCP}$, and CPER with frequency for the RCP filter ($F_5$). Notably, the device exhibits a high CPER within the operational bandwidth of 19 cm$\textsuperscript{-1}$, ranging from 684 cm$\textsuperscript{-1}$ to 703 cm$\textsuperscript{-1}$, where the CPER is greater than 20 dB, reaching a peak of 34 dB at 694 cm$\textsuperscript{-1}$, which is sufficiently large as compared to other existing RCP filters in mid-IR range. Within this range, the transmission of RCP light is greater than  49\%, reaching a peak of 51\% at  691 cm$\textsuperscript{-1}$. Further, to obtain the polarization state of transmitted light, the phase difference ($\delta$) and angle of the polarization state ($\phi$) are calculated using the relation\cite{Yang2017,trager2012springer}:

\begin{equation}
    \\ \delta = \delta_x - \delta_y
\end{equation}

\begin{equation}
        \tan(2\phi) = \frac{-2\cdot M_x\cdot M_y\cdot\cos{\delta}}{{M_x}^2 - {M_y}^2}
\end{equation}

Here, $\delta$ represents the phase difference between $x$- and $y$- components of transmitted light, $M_x$ and $M_y$ are the absolute values of transmission coefficients for $x-$ and $y-$ components of transmitted light. Fig. 2(c) shows phase difference $\delta$ ranges from -0.44$^{\circ}$ to -0.36$^{\circ}$, in the operating bandwidth. This minute variation signifies that the transmitted light is linearly polarized, with an angle of polarization state $\phi$ varying slightly about 39.02$^{\circ}$ in the entire operating bandwidth. Similarly, the achieved maximum CPER is -34 dB for filter $F_6$ (LCP filter) at 694 cm$\textsuperscript{-1}$, with a maximum $T_{LCP}$ of 51\%. (Fig. 2(d)) Further, within the operating frequency range, the CPER remained below -20 dB, and the transmission of LCP light exceeded 49\%.  The phase difference remains constant at around 179$^{\circ}$, while $\phi$ of the transmitted light varies slightly about -39$^{\circ}$, as represented in Fig. 2(e). These observations highlight that filters F5 and F6 perform exceptionally well in this frequency range with high CPER and transmittance. Further, for both these filters, $T_{RCP}$, $T_{LCP}$, $\delta$, and  $\phi$ of transmitted light from these vdW thin films are also verified numerically using the finite element method. This is shown as scatter plots in Fig. 2, which agrees well with our TMM-based theoretical calculations.

The next step involves the validation of LP filters, labelled as $F_1$ to $F_4$. Fig. 3(a) shows the frequency response of filter $F_1$, in which  LPER consistently exceeds 39 dB within the specified operating frequency range of 684 cm$\textsuperscript{-1}$ to 703 cm$\textsuperscript{-1}$, meeting the evaluation criterion. Significantly, over this bandwidth, the transmission of $x$-polarized light maintains a stable level of approximately 55\%, while the transmission of $y$-polarized light consistently remains below 0.005\%. Filters $F_2$ to $F_4$, represented in Fig. 3(b)-(d), similarly exhibit high LPER and transmission.  These filters allow the transmission of incident light linearly polarized at angles 90$^{\circ}$ ($y$-), 45$^{\circ}$, and -45$^{\circ}$, respectively, effectively blocking light polarized perpendicular to these specified directions. Further, transmittance and LPER for all these filters are verified numerically, shown as scatter plots in Fig. 3, which agrees well with our theoretical calculations. This observed behaviour underscores the precision of the filters in selectively allowing transmission based on specific linear polarization states, aligning with their designated orientations. 

Next, we present the characterization of the designed Stokes parameters for various states of polarization, including linearly polarized light along the $x$- axis, $y$- axis, 45$^{\circ}$, -45$^{\circ}$, RCP and LCP light, as shown in Fig. 4(a).  The analysis was conducted at the operating frequency 694 cm$\textsuperscript{-1}$. Stokes parameters are calculated using the Muller Matrix-based method (Eq. 4).\cite{bai2021highly} Here ${I^{F_{i}}(\omega)}$ is the transmitted intensity from the $i^{th}$ filter and  ${M_{0j}}^{F_{i}}$ (Eq. 5) is the first row of muller matrix of  the  $i^{th}$ filter. The Muller matrix elements were obtained by transforming the Jones matrix  as per Jones-Muller transformation \cite{hunte2008jones}. The Stokes parameters are obtained by solving Eq. 4 via least square fitting.  Fig. 4(b) illustrates the estimated Stokes parameters on a Poincaré sphere for an incident elliptically polarized light with equal $x$- and $y$- components of the electric field, spanning a phase difference ranging from 0$^{\circ}$ to 360$^{\circ}$. The estimated Stokes parameters (blue dots) overlap with the theoretical values (red dots) obtained using the theoretical Muller matrix elements, indicating negligible error in the estimation. Fig. 4(c) details the individual estimated Stokes parameters $S_1$-$S_3$ for an elliptically polarized light as a function of the phase difference between the $x$- and $y$-components.  It can be observed that the Stokes parameters for all the phase differences are situated entirely in the $S_2$-$S_3$ plane, with a value of zero for $S_1$, as the incident light has equal $x$- and $y$-components of the electric field, which is also evident from Fig. 4(b) where all the points lie in this plane. Additionally, when the phase difference is either 0$^{\circ}$ or 360$^{\circ}$, we expect $S_2$ to be one and $S_3$ to be 0 since the incident light is linearly polarized and oriented at 45$^{\circ}$. Similarly, for a phase difference of 180$^{\circ}$, $S_2$ and $S_3$ are expected to be -1 and 0, respectively. The maximum absolute error observed in the estimated Stokes parameters is almost zero for all three cases. These findings emphasize the accuracy and precision of the estimated Stokes parameters, showcasing minimal deviation from the theoretical values.

\begin{equation}
\begin{bmatrix}
    {I^{F_{1}}(\omega)}  \\
    {I^{F_{2}}(\omega)}  \\
    {I^{F_{3}}(\omega)}  \\
    {I^{F_{4}}(\omega)}  \\
    {I^{F_{5}}(\omega)}  \\
    {I^{F_{6}}(\omega)}  
\end{bmatrix}
= A_{6\times4}(\omega).I_{o}
\begin{bmatrix}
    {S_{0}(\omega)}  \\
    {S_{1}(\omega)}  \\
    {S_{2}(\omega)}  \\
    {S_{3}(\omega)} 
\end{bmatrix}
\end{equation}

\begin{equation}
A_{6\times4}(\omega) = 
\begin{bmatrix}
    {M_{00}}^{F_{1}}  &  {M_{01}}^{F_{1}}  &  {M_{02}}^{F_{1}} & {M_{03}}^{F_{1}}      \\
    {M_{00}}^{F_{2}}  &  {M_{01}}^{F_{2}}  &  {M_{02}}^{F_{2}} & {M_{03}}^{F_{2}}      \\
    {M_{00}}^{F_{3}}  &  {M_{01}}^{F_{3}}  &  {M_{02}}^{F_{3}} & {M_{03}}^{F_{3}}      \\
    {M_{00}}^{F_{4}}  &  {M_{01}}^{F_{4}}  &  {M_{02}}^{F_{4}} & {M_{03}}^{F_{4}}      \\
    {M_{00}}^{F_{5}}  &  {M_{01}}^{F_{5}}  &  {M_{02}}^{F_{5}} & {M_{03}}^{F_{5}}      \\
    {M_{00}}^{F_{6}}  &  {M_{01}}^{F_{6}}  &  {M_{02}}^{F_{6}} & {M_{03}}^{F_{6}}     
\end{bmatrix}
\end{equation}


\begin{figure}[h]
\centering
\includegraphics[width=\columnwidth]{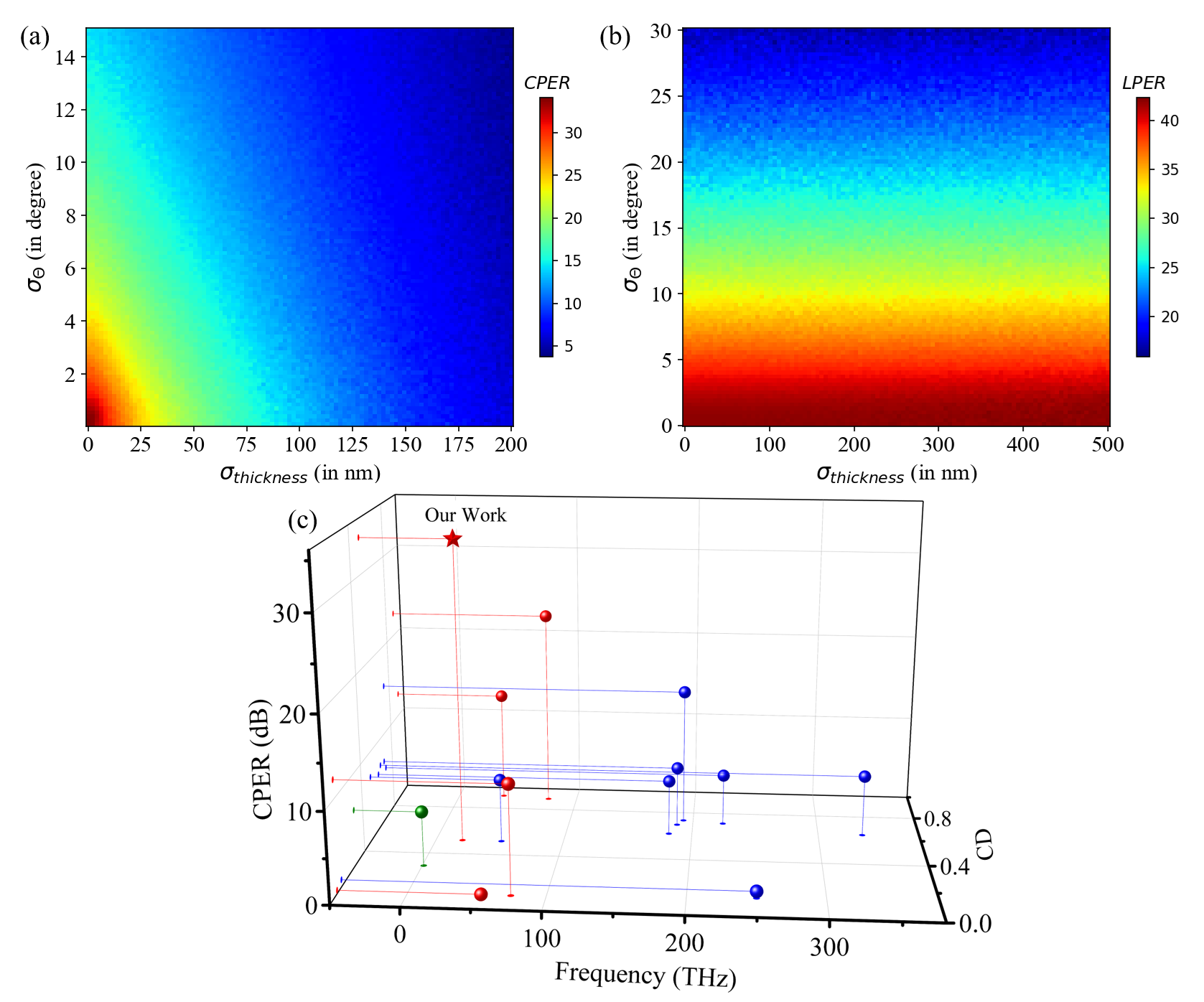}
\caption{\textbf{Fabrication tolerance metrics of the proposed polarimeter:} Variation of (a) CPER and (b) LPER, plotted at the operating frequency 694 cm$\textsuperscript{-1}$ as a function of the standard deviation of layer thickness and twist angle. This refers to the standard deviation of each $\alpha$-MoO$\textsubscript{3}$ layer present in the filters. (c) shows the comparison of CPER and CD with other existing metamaterial designs based CP filters in the IR frequency range.\cite{kuhner2023unlocking,gansel2009gold,frank2013large,bai2021highly,zhang2018nanoimprinted,ouyang2018near,bai2019chip,zhou2012terahertz,hu2017all,li2015circularly,basiri2019nature,liang2021full}  The frequency range is color-coded; blue for near-IR, red for mid-IR, and olive for far-IR. The performance of the proposed filter is represented by a star. The CPER and CD is defined as $10log(T_{RCP}/T_{LCP})$ and  $(T_{RCP}-T_{LCP})$, respectively.}
\end{figure}

To ensure the optimal performance of our polarimeter, it is crucial to evaluate the effects of any potential imperfections in its fabrication. This is particularly important given the complexity of the device's design. Therefore, we focus on assessing variations in the filter's key geometric parameters that could impact its performance. We are assessing two critical parameters: the thickness and twist angle of individual films within the filters, which could affect the filter's efficiency in polarization detection. To comprehensively evaluate the impact of these variations, we employ random sampling from a Multidimensional Gaussian distribution\cite{kumar2023universal} with mean at optimal CPER and LPER design parameters and varying standard deviation (SD). Fig. 5(a)-(b) shows the variation of CPER and LPER  at the operating frequency 694 cm$\textsuperscript{-1}$ with varying SD in the fabrication process. This was calculated by sampling out 1000 points from a Gaussian distribution with the same SD for all films in a filter and averaging out the obtained values. Adhering to our predefined evaluation criteria, ensuring CPER remains above 20 dB, we determined the maximum permissible SD for the film's thickness and twist angle, which is up to 100 nm in each film thickness and 6$^{\circ}$ twist angle could be tolerated without compromising the device performance. This indicates that despite fabrication imperfections, the system can maintain effective polarization detection within the specified range of imperfections.  Further, we compare our work with existing CP filters in the IR range. We plotted CPER and Circular Dichroism (CD) with respect to filter's operating frequency range in Fig. 5(c). Most CP filters use lithography for fabrication and have lower extinction ratios, where as our device (depicted by a star) have high CPER and moderate CD is based on non-lithography approach. \par

\section*{Conclusion}

In this study, we present the design of a polarimeter using the hyperbolic properties of $\alpha$-MoO$\textsubscript{3}$ thin films for mid-IR light detection, avoiding the need for complex lithography techniques by leveraging its inherent in-plane birefringence. The device, founded on a spatial division measurement scheme and featuring linear and circular polarizers, attains an extinction ratio exceeding 30 dB alongside a transmittance surpassing 50\% within the designated operating frequency region, demonstrating a superior performance compared to existing devices in this spectral range. The absence of lithography requirements and the ability to withstand potential fabrication imperfections highlight the practical feasibility of the proposed polarimeter. This advancement represents a significant step forward in mid-IR polarimetric detection, using the characteristics of natural hyperbolic vdW crystals for on-chip integration and a cost-effective alternative to existing mid-IR optical devices. Furthermore, its ability to seamlessly integrate with various mid-IR photodetectors and imaging arrays expands the range of potential applications in this spectral domain. Our innovative approach represents a significant step forward in mid-IR polarimetry, indicating advancement in optical instrumentation and sensing technologies.


\section*{Acknowledgement} N.R.S. acknowledges the Council of Scientific \& Industrial Research (CSIR) fellowship No: 09/087(0997)/2019-EMR-I. B.K. acknowledges support from the Prime Minister’s Research Fellowship (PMRF), Government of India. A.K. acknowledges funding from the Department of Science and Technology grants numbers CRG/2022/001170. We thank Dr Saurabh Dixit (Vanderbilt University), Dr Mihir Kumar Sahoo and Ikshvaku Shyam (LOQM, Department of Physics, IIT Bombay) for their suggestions to improvise the article.

\section*{Data availability}
\noindent The datasets generated during the current study are available from the corresponding author upon reasonable request.

\section*{References}

\bibliography{ref.bib}
\end{document}